\documentclass[runningheads]{cl2emult}

\usepackage{makeidx}  
\usepackage{graphicx} 
\usepackage{subeqnar} 
\usepackage{multicol} 
\usepackage{cropmark} 
\usepackage{physbb}   
\makeindex            

\def\ros{{\sl ROSAT }}
\def\asca{{\sl ASCA }}
\def\cobe{{\sl COBE }}
\def\scuba{{\sl SCUBA }}
\def\arcsec{\hbox{$^{\prime\prime}$}}
\def\approxlt{\mathrel{\hbox{\rlap{\lower.55ex \hbox {$\sim$}}
        \kern-.3em \raise.4ex \hbox{$<$}}}}
\def\approxgt{\mathrel{\hbox{\rlap{\lower.55ex \hbox {$\sim$}}
        \kern-.3em \raise.4ex \hbox{$>$}}}}

\begin{document}
\title*{Formation of Massive Black Holes\protect\newline in
Early Mergers ?}
\toctitle{Formation of Massive Black Holes\protect\newline in
Early Mergers}
%
%
\titlerunning{Formation of Massive Black Holes}
%
\author{Hartmut Schulz\inst{1}
\and Stefanie Komossa\inst{2}}
\authorrunning{Hartmut Schulz and Stefanie Komossa}
\institute{Astron. Inst. Ruhr University, D-44780 Bochum, Germany
\and MPI f. extraterrestrische Physik, PF 1603, D-85740 Garching, Germany}
\maketitle              
%

\vspace*{-7.0cm}
\begin{verbatim}
  Contribution to the proceedings of the workshop on
  `Trends in Astrophysics and Cosmology'
  (Physikzentrum Bad Honnef, August 24-28, 1998);
  to appear in  Lecture Notes in Physics  (W. Kundt, C. v.d.Bruck, eds) 
\end{verbatim}

\vspace*{4.5cm}

\begin{abstract}
We review theoretical and observational arguments favoring
a scenario in which a typical massive black hole (MBH) is formed
in the merger core of colliding disk systems at high $z$ during 
the build-up of a spheroid. Low-mass ($\sim 10^{5-6} M_{\odot}$) {\em 
seed} black holes
are assumed to have been formed earlier. The most massive black 
holes giving rise to the
most luminous active galactic nuclei (AGN) in active phases are 
expected to grow in violent mergers of large disk-plus-bulge systems that 
lead
to giant elliptical galaxies, the typical hosts of 
radio galaxies and quasars. We consider current ultraluminous infrared 
galaxies (ULIRG) as 
systems closely resembling the predecessors of early formed massive 
ellipticals.
As an example, we discuss the evidence for an AGN in the merging system 
NGC\,6240
which we advertize as a prototypical active ULIRG, obscured
in X-rays by a high column density absorber. 
\end{abstract}

\section{Formation of spheroids and compact cores}
There is ubiquitous evidence for the presence of massive 
(and mostly ``dark'') {\em compact} cores with $M = 10^{6 - 10} M_{\odot}$
in the nuclei of nearby galaxies ~\cite{korm}. A few of these objects, 
like
NGC\,4258~\cite{miyoshi} or the Galactic center~\cite{eckart}, are
generally considered to be nearly inevitably explained by MBHs although a
`burning disk' (BD)(\cite{kundt}, and Kundt, this conference) has been 
proposed as an alternative. 
Since convincingly detailed models for both the formation of a MBH or a 
BD are still lacking,
a clear-cut decision cannot be made
without further arguments. One argument relies on the observationally
deduced mass-energy conversion efficiency that will be briefly discussed 
in
Sect.\,1.3 below. Other common, albeit not fully compelling, arguments in 
favor of MBHs are based on variability
studies and apparent superluminal expansion~\cite{bland}. 

Below we simply follow the standard practice to identify the detected 
{\em massive compact
objects} with MBHs and discuss theoretical and observational arguments 
concerning
their formation. A large part of the discussion
can be applied to any massive compact object which is potentially able to 
promote the
non-stellar activity observed in AGNs. 

The most general prerequisites for the formation of a MBH are
(i) the presence of enough material to be compressed;
(ii) an efficient mechanism to cool the material during collapse in order 
to remove
the released gravitational energy which could otherwise inhibit the
final collapse;
(iii) an efficient mechanism to transfer angular momentum outwards
so that the inner parts of the forming disk will be able to continue
accreting material from the outer parts. Formation of
a `true' MBH means that all barriers have been overcome, also the
possible barrier put up by commencing nuclear burning.
To a certain extent, conditions (i) -- (iii)
have to be fulfilled for the formation of stars, galaxies or BDs as well.

In galaxy formation one distinguishes between {\em dissipationless} 
and {\em dissipative} collapse. A {\em collisionless} system of particles,
e.g. a stellar system or weakly interacting dark matter, can collapse
without dissipation whereas common baryonic gas is able to {\em cool}, 
i.e. lose
energy that has been gained during gravitational contraction. Domains in 
the
temperature-density diagram provide insight into the potential behavior 
of a
protogalactic gas cloud~\cite{reesostr}. Transforming the
`cooling curves' ($t_{\rm cool} = t_{\rm dyn}$) into a 
velocity dispersion - density diagram shows which objects had been
formed in a dissipative manner~\cite{korm2}. Cores of
elliptical galaxies underwent strongly dissipative collapse.
Black hole formation requires strong dissipation.
In the standard picture, the glowing accretion disks of quasars (we 
subsume
under the term `quasars' radio loud {\em and} radio quiet objects) are
the signatures of dissipative {\em growth} of black holes. 
\subsection{Black hole formation}
MBHs are widely believed to be the prime mover of non-stellar 
activity in galactic nuclei (i.e.\ in AGN; cf.~\cite{rees}, 
~\cite{bland}). Dynamical evidence for MBHs has been found
in a number of nearby galaxies~\cite{korm}.
However, the origin of MBHs is still enigmatic. A fundamental problem
is to form a {\em seed} black hole which would subsequently
grow more easily if material can be dumped down to it. 
Massive seed black holes may have formed 
(A) either directly after structure formation had entered its
epoch of non-linear collapse or (B) later via the collapse of mature
baryonic gas clouds or stellar clusters. Evidence for
MBHs has so far been only found in galactic nuclei. Therefore scenario (A)
should be related to a pre-stage or early stage of {\em formation} of
galaxies while (B) has to take place during the {\em evolution} of
galaxies. It may be interesting that luminous AGN have so far 
been found up to redshifts of
$\sim 5$, luminous quasars show indications of metal abundances higher
than solar~\cite{hamann} and that non-AGN galaxies have already been 
detected
at $z > 6$. Also, radio surveys (which are much less prone to `hiding 
effects'
as, e.g., by dust obscuration) confirm a general decline of the comoving 
quasar space
density beyond $z \sim 3$ so that ``$z > 5$ quasars must be 
rare''~\cite{shaver}.
If intergalactic hydrogen gas were neutral at these redshifts, strong 
absorption
shortward of Ly$\alpha$ would be expected in high-$z$ quasars~\cite{gunn} 
which
is not observed calling for a {\em reionization} epoch after cosmological
(re)combination. Only young stars are left as major ionizing agent so 
that we are,
altogether, led to a scenario in which an epoch
of appreciable star formation precedes MBH formation.

Tidal interaction of collapsing objects with their surroundings
leads them to acquire angular momenta which provide a
centrifugal barrier. To
circumvent this problem (the above condition (iii)) early ideas
centered on Compton drag to the electrons of the collapsing
cloud caused by the Cosmic Background Radiation~\cite{loeb1}.
This mechanism could already work at redshifts between
$\sim 200$ and the epoch of recombination provided that sufficient
reionization takes place. For the average
angular momenta expected, less exotic mechanisms like 
turbulent viscosity appear to be too slow for typical
galactic disks if they are the only agent for a
torque~\cite{loebras}. 
To overcome the centrifugal barrier
Eisenstein \& Loeb (1995,~\cite{eisen}) focused on
the {\em low-spin tail} of the distribution of angular momenta,
i.e.\ on objects that have such a low angular momentum that
a semi-relativistic disk with a short viscous time is formed which 
inevitably evolves to a black hole. 
Locations with unusually small tidal torques are expected to
be present, and Eisenstein and Loeb 
estimate that the low-spin objects should be 
abundant enough to produce $\sim 10^6 M_{\odot}$ seed black holes 
beyond $z \sim 10$ with a comoving density  comparable to that of
bright galaxies. 

The growth of black holes or even the formation of MBHs
via scenario (B) can be triggered by non-axisymmetric perturbations 
induced by bars (for a good review cf. Larson 1994~\cite{lars2})
or by tidal interactions (e.g.~\cite{barnes}).
While the usual global low-resolution 
numerical calculations verify these mechanisms to be capable to
dump down material into the innermost central kiloparsec of a galaxy
it is less clear how to reach smaller scales.
Bekki (1995~\cite{bekki}) followed the
dynamical evolution of two merging galactic cores and found
that gas can be transferred efficiently to the innermost
50 pc if the cores are rather compact and contain excessively
large amounts of gas. Inside 50 pc a MBH exceeding $\sim 10^8 M_{\odot}$
would dominate the dynamics. Scenario (B), i.e.\ the actual formation
of a MBH, would however require
to bridge the scale down to a Schwarzschild radius against
a remaining centrifugal barrier and the tendency of the gas
to fragmentation. 

A different version of scenario (B) would be to form a MBH by a
collapse of a system of more compact objects, e.g.\ a dense stellar 
cluster,
rather than from diffuse gas. This channel was one of those conjectured
for AGNs by Rees in 1984~\cite{rees} and earlier. More details about the 
growth of a
MBH were studied by David et al. (1987~\cite{david}).
A scenario of the subsequent evolution as
influenced by the circumnuclear stellar population was described
by Norman \& Scoville (1988~\cite{norman}).
Too dense stellar clusters have a low lifetime due to
frequent stellar collisions, and it is actually this argument
to infer the presence of MBHs rather than compact stellar
clusters in the nuclei of NGC\,4258~\cite{maoz}
and our Galaxy~\cite{eckart}. 

To summarize: Very early AGNs ($z > 5$) seem to be rare. If there is no
strong agent inhibiting accretion at that time, this also applies to MBHs.
Small black holes with $M \sim 10^{5-6} M_{\odot}$ may be present at
that time because they would give rise to low-luminosity AGN likely
to be outshone by star-forming sources. These small black holes
are welcome to form the {\em seeds} for subsequent growth to MBHs
which we shall discuss below. The {\em `seedless' birth} of MBHs in
present-day galaxies would be difficult from diffuse gas, but can be
imagined as the final result of star-cluster evolution provided that
a sufficiently dense cluster can be produced.  
\subsection{Formation of galaxies}
AGNs exist in spiral galaxies (most Seyfert galaxies are spirals) but
the most luminous activity and, via Eddington-luminosity 
arguments, indicative of the most massive
black holes is commonly found in ellipticals (quasars and classical
radio galaxies)~\cite{boyce}. Evidence for MBHs in nearby galaxies has 
been detected
mostly in weakly active spirals, interpreted as `dormant' black holes
starving of fuel or surrounded by an 
`advective'~\cite{abram}~\cite{naray} accretion disk. 
This suggests that MBHs are more frequent than clear-cut AGNs. 

Although selection effects are present, the mass of the MBH in nearby 
galaxies
appears to be correlated with the mass or $B$-magnitude or core radio 
power
of the spheroidal component (either bulge or the elliptical) 
(e.g.~\cite{franc}).
Relationships of AGN features with bulge properties have long been noted 
in
Seyfert galaxies~\cite{whittle}. Hence, we take it as a likely
working hypothesis that MBHs are related to the formation history of a 
spheroid.

Recent statistics on the morphology of bulges has unveiled a variety of
asymmetries that suggest that these components were not formed via a 
single
collapse and have relaxed since then. Rather a picture in terms of
gravitational interactions emerges in which the least violent ones lead to
bar formation and the subsequent development of box and peanut bulges 
while
stronger interactions lead to thick boxy 
bulges~\cite{dettmar},~\cite{l"utt}. 

On the other hand, classical textbook arguments emphasize that ellipticals
contain an old, relatively homogeneous, stellar population.
Hence, it appeared that ellipticals had been formed `at one stroke', by a 
single
collapse of low-angular momentum material followed by `violent relaxation'
and subsequent essentially passive evolution~\cite{lars1}.

Within the last decade, more and more data and calculations have 
supported 
the so-called `merger hypothesis' which implies that many ellipticals 
formed from 
the collision of disk objects. The occasional occurrence of faint shells,
tidal tails or ripples~\cite{schweiz},~\cite{seitz} and systems of young 
globular 
clusters~\cite{schweiz2} provide observational support for 
this hypothesis. Further confirmation is lent by the high stellar velocity
dispersion that has been measured in a few nearly completed mergers.
However, this does {\em not} necessarily mean that {\em all} ellipticals 
evolved 
from mergers. 

In clusters, less violent interactions, e.g. tidal stripping,
are likely to transform disk galaxies into S0s and `disky' 
ellipticals~\cite{saglia}.
Here, the
central giant elliptical (or cD) is the most likely to have evolved
via merging.
This is the strongest X-ray source and often a radio galaxy, which is in 
the
standard AGN theory a clear indication for the presence of a MBH. 
In general, luminous `boxy' ellipticals show the best observational 
clues for a merger history~\cite{bendersur}.

The scenario cannot be too simple because metal abundance 
ratios {\em rule out} that {\em massive ellipticals} could
be the product of dissipationless mergers of {\em present-day} spirals. To
explain their Mg/Fe-ratios a preponderance of type II supernovae in a
star formation time scale of $\sim 1$ Gyr is required~\cite{bender}. If 
this were
only confined to the core this might be explained by violent starburst
activity in the central molecular gas accumulated during the merging 
event (see below).
However, the observers see these abundance ratios as a global phenomenon 
of the galaxies
so that we can only maintain the merger scenario if these early merging 
spirals
differ significantly from current nearby spirals. 
   
Numerical simulations of encounters between massive disk galaxies are
able to reproduce the formation of an elliptical. A fine example
is given in~\cite{barnes} where a collision between
two milky-way type galaxies leads within $1.5\,10^9$ yrs to the 
deposition of 60\% of all
the diffuse gas of the two galaxies in the central kiloparsec
of the merger. The stars acquire a distribution typical for an elliptical.
The gas could reach the center because conditions (ii) (strong radiative
cooling) and (iii) (strong gravitational torques) were fulfilled.
If the cooling were switched off the released gravitational energy would 
have
heated the gas so strongly that a giant X-ray bubble of more than 40 kpc
radial extent would have been produced (Fig. 13 in~\cite{barnes}). We
speculate that incomplete cooling could have led to the
`seed bubbles' of current extended X-ray sources around ellipticals.

The spatial resolution of the encounter simulations is usually not good 
enough to
predict the future of the centrally collected gas. It is expected
that the gas contracts further and forms fragments so that a burst of
star formation will occur. After this stage, either a
MBH may be formed (see Sect.\ 1.1) or a seed
black hole might be fed and become more massive.

In any case, the so-called {\em ultraluminous infrared galaxies} (ULIRG; 
galaxies
with a total IR luminosity exceeding $10^{12} L_{\odot}$)
exhibit the features theoretically expected: usually 
they appear to be mergers~\cite{zou} and
they contain about $10^{10} M_{\odot}$ of molecular gas within the central
$10^{2-3}$ pc 
(e.g.~\cite{solomon}).
These objects have a space density in the nearby universe 
comparable to that of `local' quasars of the 
same $L_{\rm bol}$~\cite{sand},~\cite{kimsand}
suggesting a link to luminous AGNs.

Radio galaxies are of special interest
because, in the standard picture, they are safe to contain an AGN and a 
MBH. At $z > 3$,
their rest-frame optical morphology reveals several 10-kpc components, 
apparently dominated
by recent star formation, and partially aligned with the radio 
structures~\cite{vanbreu}.
Hence, it appears that at $z > 3$  the massive ellipticals form 
hierarchically.
At these high redshifts radio galaxies seem to evolve into more massive 
systems
than radio-quiet objects. 

Putting all the observational evidence together, present-day 
galaxies appear to have formed and
evolved during an extended merging and star formation history,
with massive ellipticals usually having completed the more violent parts 
of their history earlier,
in particular the formation of their old stellar populations. 
This rough picture resembles the more detailed theoretical
scenarios of hierarchical galaxy formation in CDM halos 
pioneered by White and Rees~\cite{white} and subsequently advanced,
among others, by the
Durham and Garching groups~\cite{baugh}. 
Seed MBHs are formed early (at high $z$) and the observed correlations 
suggest
that they grow roughly in proportion to the
spheroidal component. Whether slow `adiabatic' growth in a relaxed 
galaxy~\cite{young}
or galaxy formation around a pre-existing MBH~\cite{stiav},
takes place cannot yet be decided because both scenarios lead to similar
end states as was pointed out by van der Marel~\cite{marel}.

How much time is available for the formation of spheroids and active 
cores?
In Table 1 we give ages as a function of $z$ for two types of
unaccelerated
universes: The flat Einstein--de Sitter
world model with $q_0 = 0.5$ and, to get large time scales, the limiting 
open model 
with $q_0 = 0$.
\begin{table}
\caption{Age of the universe since big bang (for
$H_0 = 50$ km s$^{-1}$ Mpc$^{-1}$)}
\begin{center}
\renewcommand{\arraystretch}{1.4}
\setlength\tabcolsep{15pt}
\begin{tabular}{@{}lll} 
\hline\noalign{\smallskip}
Redshift $z$ & for $q_0 = 0.5$  & for $q_0 = 0$  \\
 & (years) & (years)  \\
\noalign{\smallskip}
\hline
\noalign{\smallskip}
$0$ & $13.0\,10^9$ & $19.6\,10^9$ \\
$1$ & $4.6\,10^9$ & $9.8\,10^9$  \\
$3$ & $1.6\,10^9$ & $4.9\,10^9 $ \\
$5$ & $8.9\,10^8$ & $3.3\,10^9$ \\
$20$ & $1.4\,10^8$ & $9.3\,10^8 $ \\
$100$ & $1.3\,10^7$ & $1.9\,10^8 $ \\
$1000$ & $4.1\,10^5$ & $1.9\,10^7$ \\
\noalign{\smallskip}
\hline
\noalign{\smallskip}
\end{tabular}
\end{center}
\label{Tab1}
\end{table}
Even though we adopted a low Hubble constant, Table 1 shows that
there is hardly more time than $\sim 10^9$ yrs to form galaxies
at $z \sim 3 - 5$. To assemble a galaxy with $M > 10^{11} M_{\odot}$
in $10^9$ yrs one needs star formation rates $> 10^2 M_{\odot}$ yr$^{-1}$.
In the present-day universe only ULIRGs have such star formation rates,
again suggesting that the predecessors of early ellipticals bear
similarities to ULIRGs.

However, such time-scale arguments are weakened if
the classical standard world models (that are dominated by gravitational 
breaking)
do not apply. Accelerated
world models would have a larger age at same $H_0$
and, from high-$z$ supernovae, there is mounting evidence 
for a general acceleration (cf. Ruiz-Lapuente, this conference)
\subsection{The space density of MBHs}
The present energy density $u$ of quasar radiation amounts to
$u \sim 1.3 \, 10^{-15}$ erg cm$^{-3}$ (~\cite{soltan}, 
~\cite{chokshi},~\cite{trem})
leading via standard accretion theory to an average mass density in
black hole mass 
\begin{equation}
\rho_{\rm MBH} = \frac {u} {\epsilon c^2} \approx
2.2\,10^5 \left( \frac {0.1} {\epsilon} \right) M_{\odot} {\rm Mpc}^{-3}
\end{equation}
($\epsilon$ is the accretion energy conversion efficiency, which we 
assume to be of
the order of $\sim 0.1$). With $10^{-2}$ to
$10^{-1}$ bright galaxies per Mpc$^3$ (see, e.g. Fig. 1 in
~\cite{sand}) one obtains
$M_{\rm MBH} \sim 10^{7-8} M_{\odot}$ per bright galaxy.
In the local universe Seyferts are the most abundant AGNs. Since at most
only a few percent of the spirals are known Seyferts, most MBHs 
appear to be currently inactive (or weakly active if we count
the LINERs as AGNs), i.e. `dormant quasars', as is also indicated
by the kinematical MBH searches. Although such estimates bear 
considerable 
uncertainties it appears clear that MBHs must be a frequent
component of the cores of galaxies. If the statistics were
requiring an $\epsilon$ close to the nuclear-fusion efficiency
the black-hole scenario would become less compelling as was
pointed out by Kundt~\cite{kundt}.  

Simple arguments from accretion theory suggest that 
the {\em total active lifetime} of a quasar or AGN cannot exceed
$\sim 10^{8-9}$ yrs. For example, imagine a typical
quasar with $L_{\rm bol} = 10^{46}$ erg s$^{-1}$ which requires an
accretion rate of $\sim 1 M_{\odot}$ yr$^{-1}$ so that $10^8 M_{\odot}$
will be built up within $10^8$ yrs. Note, that the Eddington luminosity of
$10^8 M_{\odot}$ is just $L_{\rm Edd}=10^{46}$ erg s$^{-1}$ and that 
actual
luminosities are slightly above $L_{\rm Edd}$. The argument can be easily 
generalized (Sect.\, 4.3) and holds as long as the accretion rate is not 
far below
`Eddington'. The fraction AGN-lifetime/Hubble time fits to the fraction 
of AGNs
among galaxies.
\section{Extended X-ray sources around ellipticals}
Up to a few keV the total X-ray luminosities of bright `normal' spiral 
galaxies are below
$\sim 10^{40 - 41}$ erg/s and the radiation can be traced back to a number
of discrete sources inside the galaxies as was verified in nearby 
galaxies~\cite{fabbia}. 
In this case the increase of $L_{\rm X}$ with the total blue luminosity 
$L_B$ of the galaxies
is relatively flat.
\begin{figure}
\begin{center}
\includegraphics[width=0.8\textwidth]{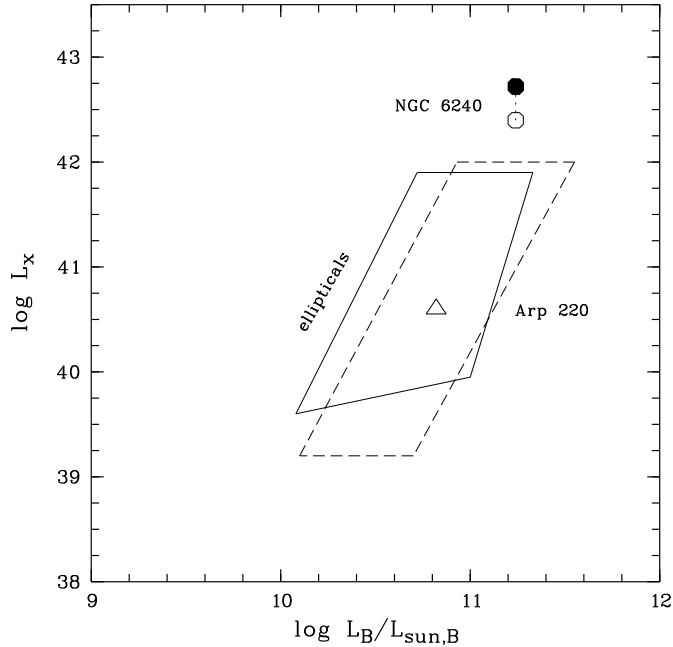}
\end{center}
\caption[]{The boxes indicate the correlation between $L_{\rm X}$ and 
$L_{\rm blue}$
for elliptical galaxies (solid line; from ref.~\cite{caniz};
dashed: from ref.~\cite{brown}). The open circle gives the
minimum observed ($0.1-2.4$ keV) X-ray luminosity of NGC\,6240, the
filled circle represents the two-component model 
adopted in~\cite{skbb} (last row of their Table 2)}
\label{eps1}
\end{figure}

On the other hand, ellipticals are generally surrounded by
X-ray bright extended sources with $L_{\rm X} \sim 10^{39 - 42}$ erg/s
that are relatively well established {\em thermal} sources in luminous 
cases. 
$L_{\rm X}$ shows a steep increase with $L_B$ (Fig. 1). For a few cluster 
ellipticals $L_{\rm X}$
of the extended source exceeds $10^{42}$ erg/s which is, however, 
attributed to the
`contamination' by the hot intergalactic medium of the cluster.
In all other closely checked cases, X-ray emission of galaxies
with X-luminosities above $10^{42}$ erg/s appears to be
related to AGNs with the most luminous sources naturally arising in 
quasars.

Taking the extended sources around ellipticals as
a quasistationary thermal X-ray bubble one obtains a typical temperature
$T \ge 7\,10^6$ K, size of $\sim 50$ kpc, mass of the hot gas of $10^{10} 
M_{\odot}$
and a mass of $\sim 10^{12} M_{\odot}$ necessary to gravitationally bind 
the bubble.
The cooling time will exceed $10^9$ yrs for pressures $n T \sim 10^4$ K 
cm$^{-3}$.
While the cooling is mostly radiative via bremsstrahlung, recombination 
and
line transitions, the usually considered heating sources are gain of 
gravitational energy,
supernova explosions and stellar-velocity dispersion ($\sigma_*$) induced 
shocks of the gas released by stellar mass loss. The latter
mechanism is supported by the above quoted
temperature and the measured $\sigma_*$ because it fits to the virial 
temperature of the stars
$T_{\rm vir} = (\mu m_p/k) \sigma_*^2 \sim 7\,10^6 (\sigma_*/300 {\rm 
km/s})^2$. 
However, there is a debate in the literature whether detailed 
hydrodynamical models
are able to match this simple model (e.g.~\cite{mathews}).
In Sect.\ 1.2 we speculated about a possible relationship of the extended 
X-ray
source to the formation history of the elliptical and its central MBH,
an idea that will be further discussed elsewhere.
\section{Early ULIRGs -- pre-stages of ellipticals and MBH cores?}
Recently, careful reduction procedures applied to the \cobe data base 
led to a reliable detection
of the apparently cosmological FIR background radiation between $140 \mu 
m$ and $5000 \mu m$ 
and its spectral shape~\cite{dwek},~\cite{fixsen}.
Modelled by direct stellar light plus dust-reprocessed stellar radiation,
notably from star-forming regions, the \cobe data
require about twice the star formation rate at $z = 1.5$ as that inferred 
from
optical-UV observations. The `added' FIR emission is likely to arise
in dust-enshrouded galaxies or star-forming regions and is consistent
with being represented by ULIRG-type spectra. A number of such `ULIRGs at 
high $z$'
(which we like to call {\em early mergers})
with typical $L \sim 3\,10^{12}$ erg/s appear to have been detected 
with the \scuba sub-mm array on the 15m James Clerk Maxwell 
Telescope~\cite{barger},~\cite{lilly}.
Although the evidence for each case rests on probabilistic 
identifications and redshifts derived 
at optical/NIR wavelengths the result appears to be statistically 
convincing.
Some disturbed morphologies strengthen the identification with early 
ULIRGs.
Lilly et al.~\cite{lilly} believe that ``these sources, producing at 
least 10\% of all stars
in the universe with star-formation rates of order $300 M_{\odot}$ 
yr$^{-1}$, are
plausibly identified with galaxies forming the bulk of their
metal-rich spheroidal component stars.''

However, the FIR radiation is emission reprocessed by dust so that the 
true 
heating source at lower wavelengths is concealed. For simplicity it is 
being
attributed solely to star formation. In local ULIRGs there is an ongoing 
debate whether
AGNs do partly contribute to the FIR power~\cite{sand}.

Observationally it appears that the power of LIRGs (luminous infrared 
galaxies, 
$L_{\rm FIR} < 10^{12}\,L_{\odot}$)
can be explained by predominant star-formation 
and that the evidence for an AGN contribution increases with FIR 
luminosity
(e.g.\cite{shier},~\cite{lutz},~\cite{rigo}; see 
\cite{sand} and~\cite{genzel} for recent reviews). In particular,
essentially all of the HyLIRGs (hyperluminous infrared galaxies,
$L_{\rm FIR} \gg 10^{12}\,L_{\odot}$)
seem to contain quasars~\cite{hines}. 
In the `transition region' around 
$L_{\rm FIR} \simeq 10^{12}\,L_{\odot}$
it then requires a careful object-by-object analysis to find out the 
major power source.    

As in local mergers, MBHs might
have grown in early mergers as well. We shall return to this point in the
concluding section after having described the case for an AGN in a local
ULIRG which we consider as a prototype.
\section{An obscured AGN in the ULIRG NGC\,6240?}
With a redshift $z=0.024$ and a far-infrared
luminosity of $\sim 10^{12} L_{\odot}$ (cf.~\cite{wright}), 
NGC\,6240 is one of the nearest members of the class of ULIRGs (although
due to changes of the methods to integrate
over the wide IRAS bands and the adopted value of $H_0$, most authors
now attribute an IR luminosity $<10^{12}\,L_{\odot}$ to NGC\,6240
rendering it a LIRG instead of a ULIRG; favoring a small $H_0$ and 
considering
it more typical for the `upper class' we continue to call it 
an ULIRG). 
This object has remarkable properties: its infrared
H$_2$ 2.121$\mu$m and [FeII] 1.644$\mu$m line luminosities
and the ratio of H$_2$ to bolometric luminosities are the
largest currently known~\cite{vdW}; 
its apparent tidal tails and loops~\cite{fosb} 
and a double nucleus~\cite{fried} complemented
by its large stellar velocity dispersion of 360 km/s
(among the highest values ever found in the center
of a galaxy,~\cite{doyon}) suggest that it is
a merging system on its way to become an elliptical. 
Like other ULIRGs, the object contains a compact, only $\sim 10^2$ pc 
large,
luminous CO(1-0) emitting core of molecular
gas~\cite{solomon}.
Within this core most of the ultimate
power source of the FIR radiation appears to be hidden.
 
Concerning NGC\,6240, at least four types of power 
sources have been suggested: 
Heating of dust by a superluminous starburst, by an AGN, 
by an old stellar population, and 
by UV radiation from molecular cloud collisions. In particular, 
previous hints for an AGN included (i) the strength of near-infrared 
recombination
lines~\cite{depoy}, 
(ii) the presence of compact bright radio cores~\cite{carral}
(iii) the discovery of a high-excitation core
in the southern nucleus with {\sl HST}~\cite{barb1},~\cite{barb2}, 
~\cite{rafa},
and the detection of the [OIV]\,25.9$\mu$m emission line with {\sl 
ISO}~\cite{lutz96}.
 
None of these arguments is watertight: (i) modified starburst models and 
special extinction situations could explain the near-infrared hydrogen 
recombination lines;
(ii) a special arrangement of young radio supernovae could account for 
the 
compact radio sources~\cite{colb}; (iii) the high-excitation core
could be due to a `young flame' of star formation in an older environment;
(iv) the high-excitation [OIV]\,25.9$\mu$m line has been found in 
some starburst galaxies as well~\cite{lutzb}. 

In this contribution we emphasize the role of X-rays as a powerful tool 
to obtain clues for both an AGN and starburst-superwind activity.   
We discuss the evidence for a hard X-ray component in the
\ros PSPC spectrum of NGC\,6240~\cite{skbb}
as well as the discovery of luminous extended
emission based on \ros HRI data~\cite{ksg} in combination with
the observation of an FeK line and a hard X-ray 
component by \asca (first reported by Mitsuda (1995,~\cite{mitsuda}). 
Altogether,
these findings strongly suggest the presence
of an obscured AGN.

Luminosities given here correspond to $H_0 = 50$ km/s/Mpc.
\subsection{The X-ray spectrum of NGC\,6240}
The \ros PSPC spectrum of NGC\,6240
could not be satisfactorily reproduced by a {\em one-component} 
fit~\cite{skbb}.
E.g., a single Raymond-Smith model requires a huge absorbing column
along the line-of-sight, the consequence being an intrinsic (absorption
corrected) luminosity of $L_{\rm X,0.1-2.4} \simeq 4\,10^{43}$ erg/s,
practically impossible to be reached in any starburst-superwind scenario
adjusted to the optical observations of NGC\,6240.
Our checks with a variety of fits confirm Fricke \& Papaderos 
(1996,~\cite{fricke}) who
had claimed 3.8\,10$^{42}$ erg/s with a thermal bremsstrahlung model.
We found an excellent estimate of a lower limit of the intrinsically 
emitted X-ray luminosity
by fitting a single black body that did not require excess absorption: 
$L_{\rm X} \approxgt 2.5\,10^{42}$ erg/s in the (0.1-2.4) keV band
(dealing with $H_0$ and other uncertainties yields a rather firm lower 
limit
$1\,10^{42}$ erg/s).
However, {\em acceptable fits} require at least
{\em two-components} one of which has to be a {\em hard} X-ray component 
that can be represented by either very hot thermal emission ($kT \simeq$ 
7 keV) or a
powerlaw. This applies for solar abundances or not too far from solar. 

Admittedly, a second component in
the \ros band could be omitted if strongly depleted metal abundances
are adopted. At the present stage of the art of 
modelling (see the discussion in Komossa \& Schulz (1998,~\cite{ks}) and 
Buote \& Fabian (1998,~\cite{buote}), we consider such a solution as 
unlikely
for NGC\,6240. We are more concerned about other changes in the models due
to future theoretical improvements.

The two-component fits yield a luminosity of the hard \ros component
of a few $10^{42}$ erg/s, at least $1\,10^{42}$ erg/s.
Due to the largeness of this luminosity we regard
the hard component as {\em ultimately} arising in an AGN, i.e. as
a scattered AGN component.  
Various fits of \asca spectra (e.g.,~\cite{mitsuda},~\cite{kii},
~\cite{iwasawa},~\cite{netzer})
reveal the extension of the hard component up to 10 keV. Hence, the 
presence 
of a rather hard and luminous component is well substantiated although
there are some clear differences in the fits of the various authors.  
The approaches differ in the description
of the soft component(s) and the amount of absorption of the
hard component. In any case, it is striking that 
the luminosity of the $(2-10)$ keV radiation 
amounts to a few $10^{42}$ erg/s as well.

As we saw above, from UV to radio wavelengths, no convincing
evidence has been found for a {\em directly
seen} AGN, but circumstantial evidence for the presence of an 
{\em obscured} AGN. The first analysis of the {\asca}spectrum of 
NGC\,6240 
already revealed a conspicuous FeK line complex with a high
equivalent width of $\sim 2$ keV typical for an AGN with some sort of
scattering geometry. The \asca spectrum resembles that of NGC\,1068, the
well studied `prototype' of a hidden AGN~\cite{ueno}. Adopting 
the same scattering geometry as in NGC\,1068 the {\em intrinsic}
X-ray luminosity should be about $\sim 10^2$ times larger than that of the
observed scattered radiation, yielding at least $\sim 10^{44}$ erg/s  for 
the
hard component which translates into $L_{\rm bol} ({\rm AGN}) \sim 
10^{45}$ erg/s.
This means that the AGN contributes an appreciable part of 
$L_{\rm FIR} \sim 4\,10^{45}$ erg/s. A similar result was obtained from
\ros data by us by guessing the scattering geometry from the optically
visible apparent H$\alpha$ cone~\cite{skbb}. Also, another type of 
scattering model,
with FeK$\alpha$ arising in a highly ionized near nuclear
`warm scatterer' (in other directions it would appear as a `warm 
absorber') 
would require a similarly luminous AGN~\cite{ksg}.
The latter model, which was suggested to explain the hard component,
bears some similarities to the one suggested by Netzer et al. 
(1998,~\cite{netzer}).

However, Netzer et al. explained the {\em whole} \asca spectrum 
in terms of scattering and this seems to be ruled by the huge extent
of the X-ray source ($ > 25$ kpc) disclosed in ref.~\cite{ksg}
with the \ros HRI. We showed that efficient
scattering can only be made with a small scatterer~\cite{skbb}.
A small scatterer and an AGN are nearly inevitable if the
X-ray variability detected in ref.~\cite{ksg} is 
confirmed. So far, the significance of the variations is only taken 
with some caution because it rests on measurements with different 
instruments.   

Summarizing, the presence of the hard X-component and the high
equivalent width FeK$\alpha$ provide excellent complementary
evidence for an obscured AGN in addition to the less compelling
indications cited in the introduction to Sect.\,4. After briefly 
commenting
on the {\em extended} X-ray emission we shall discuss the
consequences.
\subsection{Extended X-ray emission}
\begin{figure}
\begin{center}
\includegraphics[width=0.8\textwidth]{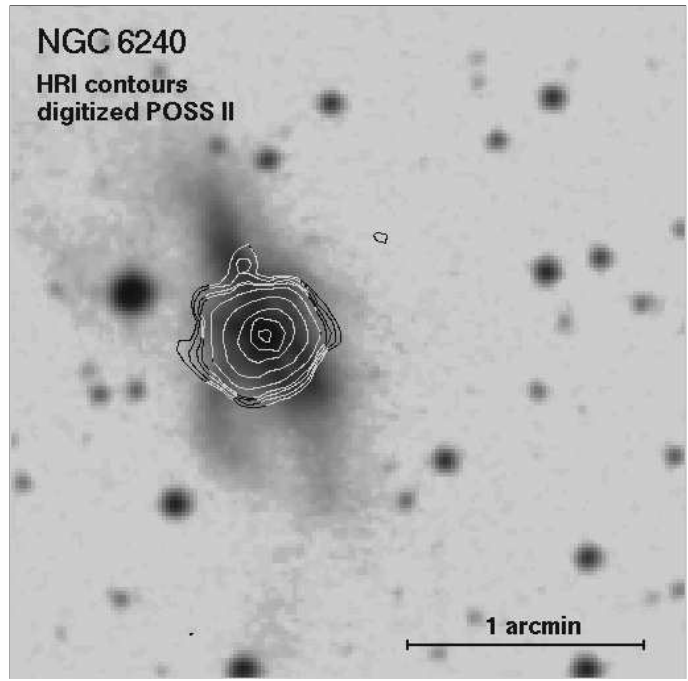}
\end{center}
\caption[]{\ros HRI X-ray contours overlaid on a Palomar Sky survey
image of NGC\,6240. The outermost contour is at $2\sigma$ above
the background of the X-ray frame}
\label{eps2}
\end{figure}

The HRI {\em images}~\cite{ksg}
reveal that part of the huge X-ray luminosity arises in a
roughly spherical source
with strong ($\ge 2\sigma$ above background) emission out to a
radius of 20\arcsec~($\sim$14 kpc; Fig. \ref{eps2}).
Hence, NGC\,6240 is the host of one of
the {\em most luminous extended} X-ray sources
in isolated galaxies (see Fig.\,1 where $L_{\rm X}$ is compared
with a sample of elliptical galaxies and Arp\,220).
Analytical estimates based on the Mac Low \& McCray (1988,~\cite{maclow}) 
models
show that the {\em extended} emission can be accounted for by 
superwind-shell interaction
from the central starburst~\cite{skbb}.

However, unless the starburst source is off-center
this scenario lets one expect a bipolar rather than
the observed near-circular symmetry of the
X-ray source. Could the X-ray bubble have been formed
due to expansion of some of the merger-induced infalling gas
because of incomplete cooling?  
Considering that NGC\,6240 is on its way to become an elliptical
we~\cite{ks99}~\cite{schutrum} mentioned a few further possibilities
like heating 
by the large velocity dispersion of $(350-360)$ km/s 
(\cite{lester},~\cite{doyon})
or that the bubble might have been caught
in the verge of experiencing its central cooling
catastrophe~\cite{ciotti}~\cite{friaca}.
Shocks induced by a cooling flow might also help to explain
the LINER-like line ratios in the central few kpc
and, with lower velocities, excite the molecular cloud complex between
the nuclei leading to the extreme H$_2$ luminosity found there~\cite{vdW}.
\subsection{Inferences on the MBH in NGC\,6240}
In the interpretation suggested above, the {\em extended} X-ray emission
is not influenced by the present AGN which could only contribute to a
{\em near-nuclear} scattered component. In Sect.\,4.1 we derived  
$L_{\rm bol}{\rm (AGN)} \sim 10^{45}$ erg/s. Assuming that this value
is close to the Eddington luminosity, 
a black-hole mass of $M_{\rm MBH} \simeq 10^7 M_{\odot}$ results. 

One might argue that the present
black hole mass could be $\gg 10^7 M_{\odot}$ while the accretion rate is
significantly below the Eddington rate. However, in the simplest
scenarios a low feeding rate is unexpected
in an object where plenty of fuel is
available, and where the infall-triggering forces are present.
We therefore consider $M_{\rm MBH} \sim 10^7 M_{\odot}$ as the
currently best estimate for NGC\,6240.

NGC\,6240 is expected
to form an elliptical galaxy of $10^{11-12} M_{\odot}$
after having completed its merging epoch~\cite{shierf}.
Since the relation~\cite{lauer} 
\begin{equation}
M_{\rm MBH} \approx 0.002\, M_{\rm gal}
\end{equation}
valid for the evolved elliptical, predicts $M_{\rm MBH} \sim 10^{8-9} 
M_{\odot}$
the black hole has still  
to grow by one or two orders of magnitude.

Equalizing a typical accretion luminosity 
$L_{\rm acc} = 10^{46} ({\rm d}M/{\rm d}t) / (1 M_{\odot}/{\rm yr})$ erg/s
with the Eddington luminosity $L_{\rm Edd} = 1.3\,10^{38} (M/M_{\odot})$ 
erg/s
we get an ``Eddington accretion rate'' ${\rm d}M/{\rm d}t = 7.7\,10^{-9} 
M$ if
$t$ is measured in years and $M$ in $M_{\odot}$. Integrating this equation
leads to exponential growth with
\begin{equation}
M = M_{\rm init}\, 10^{t/3\,10^8 {\rm yr}}
\end{equation}
Consequently, under optimal conditions the MBH could grow 
to $\sim 10^9 M_{\odot}$ within $6\,10^8$ yr although, in practice,
a longer time scale is expected.
The fuel for the growth of the MBH 
finally comes from the $10^{10} M_{\odot}$ of molecular gas residing in 
the
core of NGC\,6240~\cite{solomon}.
According to these simple budget considerations there appears to be no
difficulty in obtaining the expected mass of the black
hole in the resulting elliptical. 

However, many details of the actual astrophysical MBH-growth processes 
have still
to be worked out.
As yet,
some authors (e.g.~\cite{trem}) believe that mergers usually convert 
galactic disks 
to bulges without any corresponding change in the prior black hole mass 
while
others (like us) take it for granted that the merger event leads to rapid
growth of the black hole(s). Recently, Wang \& Biermann~\cite{yiping} 
showed
by a beautiful accretion model that the empirical relation Eq.\,2 can be
understood quite well if one takes the competetive feeding of starburst 
and MBH
into account. 
\section{Concluding discussion and outlook}
The particular merger NGC\,6240 has so far
revealed several strong clues for the presence of an AGN and may be a 
rosetta stone.
In addition to MBH evidence, it exposes a gigantic luminous extended 
X-ray source
similar to the X-ray sources of ellipticals.
Therefore we suggested ingredients beyond a standard starburst outflow 
model (Sect.\,4.2),
among those that the bubble reflects part of the energy from the infall 
history of the ISM
of this galaxy. The compressed molecular gas core provides the fuel
for the central MBH, giving rise to AGN activity. 

Has the MBH been formed during the merger or was there a black hole 
before?
The correlation~\cite{lauer} between black hole mass and mass of the 
spheroid
and the statistical consistency with MBHs being a common component
of spheroids (bulges and ellipticals) (Sect.\,1.3) suggests   
a hierarchical picture: spiral bulges contain MBHs that will be seeds for 
a bigger MBH formed in a merger that evolves into an elliptical. 
Thus it is one (or two) preexisting MBH(s) that is (are) growing in 
NGC\,6240. 

In this vein, mergers of lower-mass galaxies produce lower-mass MBHs,
while the most luminous AGNs are expected in encounters of very massive 
objects.
Because the molecular material in the central $\sim 10^2$ pc will form a 
disk,
the visibility of the AGN depends on the viewing angle and the accretion 
mode.
To satisfy eq.\ 2, about a few percent of the central gas has to be 
dumped onto 
the growing MBH, most of the rest has to be consumed by the starburst to 
lead
to a gas-poor elliptical. As long as the starburst lasts, the AGN will 
not be
the dominating luminosity source of the ULIRG.

Can this picture be transformed to much earlier epochs of the
universe when most of the massive ellipticals had been formed?
So far, Hubble deep field observations, the \cobe infrared background and 
\scuba sub-mm identifications of infrared luminous high-$z$ galaxies 
observations
broadly support a hierarchical galaxy formation picture that was
theoretically developed as a kind of cold dark matter 
structure formation~\cite{white},~\cite{kauff},~\cite{baugh}. The merging
of baryonic material occurs in cold-dark matter halos that had been
earlier assembled by merging processes as well. 

Concerning
MBH formation, sceptics argue that the peak of the number
of luminous quasars at high $z$ is hard to understand
in a hierarchical scenario because low-mass black
holes form first and MBHs will grow by the course of time. 
This simple argument is opposed by Haehnelt \& Rees~\cite{hae} 
who, in their quasar-evolution
scheme, showed that a MBH-formation efficiency (parametrized
by $M_{\rm MBH}/M_{\rm Halo}$) proportional to a high power of $(1+z)$ is
conceivable. Observationally, counts of galaxies suggests a growth of the
pair density with a power of $(1+z)$ leading to a correspondingly higher
merger rate.  

However, ``early mergers'' (i.e., at high $z$) are likely to be be quite 
different from mergers
today. The numerical studies focus on detailed Monte-Carlo treatment  
of the dark-matter component and treat the baryonic gas only in a 
schematic way.
Detailed studies of nearby mergers are important in order to learn more 
physics
to be included in the simulations of forming and evolving galaxies.

Our scheme in which MBHs are evolving during the assemblage
of the spheroidal components of galaxies has to start with some kind
of seed black holes. The Eddington
growth rate of Eq.\,3 can be taken as a crude upper limit for MBH growth.
Starting with the Eisenstein-Loeb~\cite{eisen} seeds 
of $\sim 10^6 M_{\odot}$ (Sect.\,1.1)
we need three to four 10-folding times for $M_{\rm MBH} = 10^{9-10} 
M_{\odot}$,
i.e. at least $\sim 10^9$ yrs. Such massive MBHs exist at $z \sim 5$,
the Eisenstein-Loeb seeds were formed beyond $z \sim 10$. Table 1 shows
that this growth requirement excludes an Einstein-de Sitter Universe 
($q_0 = 0.5$) and
calls for a tenuous open universe with a low value of $q_0$ unless the
universe is accelerating. A more precise statement on the allowed 
world models is not warranted without simultaneously modelling structure 
formation.

In any case, time scales are tight, hopefully also for the new telescopes
and instruments that will help us to verify or modify the picture of
formation and evolution of spheroids and massive black holes outlined 
above.

\vspace{1.7cm}
\noindent St.K. acknowledges support from the Verbundforschung under 
grant No. 50\,OR\,93065.
The {\sl ROSAT} project has been supported by the German 
Bundes\-mini\-ste\-rium
f\"ur Bildung und Wissenschaft (BMBW/DLR) and the Max-Planck-Society. \\
A preprint of this and related papers can be retrieved from our
webpage at http://www.xray.mpe.mpg.de/$\sim$skomossa/ 
%
%

\clearpage
\addcontentsline{toc}{section}{Index}
\flushbottom
\printindex

\end{document}